\documentclass[aps,prl,twocolumn]{revtex4-2}
\usepackage{amsmath}

\usepackage{newtxtext,newtxmath}
\usepackage{dcolumn}% Align table columns on decimal point
\usepackage{bm}

\newcommand{\beq}{\begin{equation}}
\newcommand{\eeq}{\end{equation}}
\newcommand{\bea}{\begin{eqnarray}}
\newcommand{\eea}{\end{eqnarray}}

\begin{document}

\title{On the Fast Electrons-Optical Modes Interaction Problem Error}

\author{Svetlana Churochkina}
\email[]{klechshevskaya@mail.ru}
\author{Dmitry Churochkin}
\email[]{churd2000@mail.ru}

 \affiliation{Department of
Physics, Saratov State University, 83 Astrakhanskaya, Saratov
410012, Russia}


%\author{}
%\thanks{}
%\affiliation{}


\date{\today}

\begin{abstract}
A recently developed relativistic quantum optics approach for a
fast electron beam-optical modes interaction problem (Optica 6,
1524--1534 (2019)) contains suggestion about a smallness of the
error induced by replacement $\Psi_{\vec k}\approx\Psi_{\vec k_0}$
for noninteracting free electron Dirac spinors at such wave vector
$\vec k$ that $\left|\vec k-\vec k_0 \right| \ll\left|\vec
k_0\right|$ with $\vec k_0$ as a peaked wave vector of a focused
electron beam. Here, based on quantum information approach we
strictly investigated a degree of coincidence of spinors
$\Psi_{\vec k}$ and $\Psi_{\vec k_0}$ by calculating fidelity for
them. The analytical expression is obtained. It is shown that the
fidelity inevitably depends on not only wave vectors $\vec k$ and
$\vec k_0$ but also on mutual orientation of $[\vec k_0,\vec k]$
and $\hat{\vec{s}}^{\dagger}\vec{\sigma}\hat{\vec{s}}$ with
$\vec{s}$ is a two-component spin unit vector and $\vec{\sigma}$
is a vector of Pauli matrices. This result can be conceived as a
chirality-like effect in the fidelity and constraints spin
arbitrariness when applying the developed relativistic quantum
optics approach to the problems of interest. A geometrical
configuration of $\vec k_0$, $\vec k-\vec k_0$, and
$\hat{\vec{s}}^{\dag}\vec\sigma\hat{\vec{s}}$ that is optimal in
terms of fidelity is established.
\end{abstract}

%\keywords{}

\maketitle

\section{Introduction}

In accordance with the model~\cite{Abajo2019}, a noninteracting
free electron Dirac spinor with relativistic momentum $\hbar\vec
k$ and energy $E_k=c\sqrt{m^2_e c^2+\hbar^2 k^2}$ is expressed as
\bea \label{spinor} \Psi_{\vec k}=A_k \left(\begin{array}{c}
                \hat{\vec{s}} \\
               B_k(\vec k\cdot\vec\sigma) \hat{\vec{s}}
              \end{array}\right),
\eea
where $\hat{\vec{s}}$ is a two-component spin unit vector
represented as a column of components $s_1$ and $s_2$, i.e.
$\left(\begin{array}{c}
                s_1 \\
               s_2
              \end{array}\right)$ , $A_k=\sqrt{\left(E_k+m_e c^2\right)/2E_k}$, $B_k=\hbar c/\left(E_k+m_e c^2\right)$, and $\vec\sigma$ is a vector of Pauli matrices:\\
             %$\sigma_1=\left(\begin{array}{c}
              %  0\quad  1 \\
               %1\quad  0
              %\end{array}\right)$,$\sigma_2=\left(\begin{array}{c}
               % 0\quad -\imath \\
               %\imath \quad 0
              %\end{array}\right)$,$\sigma_3=\left(\begin{array}{c}
               % 1\quad 0 \\
               %0\quad -1
              %\end{array}\right)$.
              \begin{equation}  \label{paulim}
        \sigma_x =
      \begin{pmatrix}
  0& 1\\
  1& 0
    \end{pmatrix}, \quad
\sigma_y =
      \begin{pmatrix}
  0& - i\\
  i& 0
    \end{pmatrix},
  \quad
\sigma_z =
      \begin{pmatrix}
  1& 0\\
  0& - 1
    \end{pmatrix}.
   \end{equation}

The problem of interest here is an electron beam with a peaked
wave vector $\vec k_0$ such that the other beam wave vectors $\vec
k$ are sufficiently close to $\vec k_0$, i.e., $\left|\vec k-\vec
k_0 \right| \ll\vec k_0$. This makes it possible to make an
approximation $\Psi_{\vec k}\approx \Psi_{\vec k_0}$ throughout
the model, assuming that the error induced by this condition would
not be crucial. Considering that the model becomes widespread
under description of different aspects of free electron
beam-matter interaction~\cite{Velasco2024,Velasco2026}, we found
it reasonable to bring a little bit different view on the problem.
Indeed, the  concept of an "error" naturally send us to a quantum
information playground where this idea is commonly
used~\cite{Nielsen2010}. Thus, we tried to apply the quantum
information meaning of error to  the analysis of the physical
problem mentioned above. In other words, what is the meaning of
smallness of an error between two Dirac spinors in terms of
quantum information? It is known that, in the quantum information
science, there is many different measures of distinguishability of
different quantum states~\cite{Nielsen2010}. In the course of
choosing the appropriate measure, we were guided by the idea of
simplicity of its calculation as well as possibility for making
clear physical conclusions on the basis of the corresponding
analytical result for the chosen measure. Following these reasons,
we believe that fidelity could be a relevant measure of interest
that can be easily adopted to the case of spinors and, as it will
be clear below, calculated analytically.
\section{Fidelity evaluation}

We firstly notice that the spinors $\Psi_{\vec k}$, $\Psi_{\vec
k_0}$ are normalized to unity that is \bea \label{norm}
\Psi^{\dag}_{\vec k}\Psi_{\vec k}=\Psi^{\dag}_{\vec k_0}\Psi_{\vec
k_0}= 1, \eea where $^{\dag}$ is the Hermitian conjugate.

Then, the fidelity $F$ for two spinors at wave vectors $\vec k$
and $\vec k_0$, representing pure quantum states, can be defined
as follows \bea \label{fidel} F=\left|\Psi^{\dag}_{\vec
k_0}\Psi_{\vec k}\right|^{2}. \eea

Hence, the condition  $\Psi_{\vec k}\approx\Psi_{\vec k_0}$ in
terms of fidelity signifies \bea \label{fidelII} F\approx 1. \eea

Thus, from the point of view of quantum information the problem is
reduced to the analysis of regimes when fidelity is close to unity
from below, say, for example, belongs to the range of \bea
\label{fidelIrange} \eta\leq F\leq 1, \eea with $\eta$ typically
more than 0.9, but less than 1, suggesting that when F is above
0.9, the error is small, in accordance with commonly accepted
quantum information numerical constraints for $F$.

Having in mind these numerical constraints, let's evaluate $F$
analytically. To this end we have to introduce a couple of known
auxiliary relations, which are useful for further calculations.

First, the fact that $\hat{\vec{s}}$ is a spin unit vector
signifies normalization to unity \bea \label{spinunit}
\hat{\vec{s}}^{\dag}\hat{\vec{s}}=|s_1|^2+|s_2|^2=1. \eea

Second, the Pauli matrices have the following
property~\cite{Sakurai}

\bea \label{pmprop}
\sigma_i\sigma_j=\delta_{ij}I+\imath\epsilon_{ijk}\sigma_k, \eea
with $i,j,k=1,2,3$, $\delta_{ij}$ and $\epsilon_{ijk}$ as the
Kronecker delta and Levi-Civita symbols, respectively, and $I$ as
a two by two unit matrix. Here, the Einstein summation convention
for indices is applied.

Now, the explicit expression for fidelity can be written as
follows \bea \label{fidelII}
\begin{array}{c}
F=\left|\left(A_{k_0} \left(
    \begin{array}{c}
                \hat{\vec{s}} \\
               B_{k_0}(\vec k_0\cdot\vec\sigma) \hat{\vec{s}},
    \end{array}
\right)\right)^{\dag}\left(A_k \left(
    \begin{array}{c}
                \hat{\vec{s}} \\
               B_k(\vec k\cdot\vec\sigma) \hat{\vec{s}},
    \end{array}
\right)\right)\right|^{2}\\
             = \left|A^{*}_{k_0}A_k \left(
    \begin{array}{c}
                \hat{\vec{s}}^{\dag}\quad
               B^{*}_{k_0}\left((\vec k_0\cdot\vec\sigma) \hat{\vec{s}}\right)^{\dag}
    \end{array}
\right)\left(
    \begin{array}{c}
                \hat{\vec{s}} \\
               B_k(\vec k\cdot\vec\sigma) \hat{\vec{s}},
    \end{array}
\right)\right|^{2}

\\
              = \left|A^{*}_{k_0}A_k \left(\hat{\vec{s}}^{\dag}\hat{\vec{s}}+B^{*}_{k_0}\left((\vec k_0\cdot\vec\sigma) \hat{\vec{s}}\right)^{\dag}B_k(\vec k\cdot\vec\sigma) \hat{\vec{s}}\right)\right|^{2}

\\
             = \left|A^{*}_{k_0}A_k \left(1+B^{*}_{k_0}B_k\hat{\vec{s}}^{\dag}(\vec k_0\cdot\vec\sigma)^{\dag}(\vec k\cdot\vec\sigma) \hat{\vec{s}}\right)\right|^{2}

\\
             = \left|A^{*}_{k_0}A_k \left(1+B^{*}_{k_0}B_k\hat{\vec{s}}^{\dag}( k_{0i}\sigma_{i})^{\dag}(k_{j}\sigma_{j}) \hat{\vec{s}}\right)\right|^{2}

\\
              =\left|A^{*}_{k_0}A_k \left(1+B^{*}_{k_0}B_k\hat{\vec{s}}^{\dag}( k^{*}_{0i}k_{j}\sigma_{i}\sigma_{j}) \hat{\vec{s}}\right)\right|^{2}

 \\
              = \left|A^{*}_{k_0}A_k \left(1+B^{*}_{k_0}B_k\hat{\vec{s}}^{\dag}( k^{*}_{0i}k_{j}(\delta_{ij}I+\imath\epsilon_{ijl}\sigma_l))
              \hat{\vec{s}}\right)\right|^{2}

  \\
              = \left|A^{*}_{k_0}A_k \left(1+B^{*}_{k_0}B_k( k^{*}_{0i}k_{j}(\delta_{ij}\hat{\vec{s}}^{\dag}I\hat{\vec{s}}+\imath\epsilon_{ijl}\hat{\vec{s}}^{\dag}\sigma_l\hat{\vec{s}})) \right)\right|^{2}

 \\
              = \left|A^{*}_{k_0}A_k \left(1+B^{*}_{k_0}B_k\left(\vec k^{*}_{0}\cdot\vec k+\imath\left[\vec k^{*}_{0},\vec k\right]
               \cdot\left(\hat{\vec{s}}^{\dag}\vec\sigma\hat{\vec{s}}\right)\right) \right)\right|^{2}.

\end{array}
\eea

Here, the first equality follows from Eqs.~(\ref{spinor}) and
(\ref{fidel}). The second equality is the result of application of
the Hermitian conjugate operation to $\Psi_{\vec k_0}$. The third
equality is a result of the row-column product. To derive the
fourth equality, Eq. (\ref{spinunit}) and the rule for taking the
Hermitian conjugate of a product of operators are used. The fifth
equality is the same as the fourth one, but with the corresponding
scalar products explicitly written in components. The sixth
equality is obtained when the Hermitian conjugate operation is
performed in one of the scalar products. The seventh equality is
derived using Eq. (\ref{pmprop}). The eighth equality is a result
of taking the wave vectors components out of the corresponding
bracket. To get the last equality, we again use Eq.
(\ref{spinunit}), as well as the definition of a vector product
through the Levi-Civita symbol.

We note that this is the key analytical result of the article. It
can be modified further if we recall that the wave vector in Eq.
(\ref{spinor}) is supposed to be with the real-valued components.
It means that \bea \label{real} \vec k^{*}=\vec k \quad
\Rightarrow \quad A^{*}_{k}=A_{k}; \quad B^{*}_{k}=B_{k}. \eea

In addition, we observe that
\bea \label{realspinvec}
\begin{array}{c}
\left(\hat{\vec{s}}^{\dag}\vec\sigma\hat{\vec{s}}\right)^{*}=
\left(\left(\hat{\vec{s}}^{\dag}\vec\sigma\hat{\vec{s}}\right)^{*}\right)^{T}=
\left(\hat{\vec{s}}^{\dag}\vec\sigma\hat{\vec{s}}\right)^{\dag}\\
= \left(\vec\sigma\hat{\vec{s}}\right)^{\dag}\hat{\vec{s}}=
\hat{\vec{s}}^{\dag}\vec\sigma\hat{\vec{s}}, \end{array}
 \eea
 therefore,
$\hat{\vec{s}}^{\dag}\vec\sigma\hat{\vec{s}}$ is a vector with
three real-valued components. To get this result we formally
treated every component (i.e., a number) of vector
$\hat{\vec{s}}^{\dag}\vec\sigma\hat{\vec{s}}$ as a one by one
matrix and used the general rules for operations  of complex
conjugation $^{*}$, transposition $^{T}$, and $^{\dag}$. Moreover,
recalling the explicit expressions for $\hat{\vec{s}}$ and
$\vec\sigma$ given above, one can easily show that
$|\hat{\vec{s}}^{\dag}\vec\sigma\hat{\vec{s}}|=1$, therefore,
$\hat{\vec{s}}^{\dag}\vec\sigma\hat{\vec{s}}$ is a unit vector.

Collecting Eqs. (\ref{fidelII}), (\ref{real}), and
(\ref{realspinvec}) together, we come to the following result
\bea\label{fidelIII}
\begin{array}{c}
F=A^{2}_{k_0}A^{2}_k\left(\left(1+B_{k_0}B_k\vec k_{0}\cdot\vec
k\right)^{2}\right.\\
\left.+\left(B_{k_0}B_k\left[\vec k_{0},\vec
k\right]\cdot\left(\hat{\vec{s}}^{\dag}\vec\sigma\hat{\vec{s}}\right)
\right)^{2}\right). \end{array} \eea
\section{Perturbative error analysis and optimal configuration}
It can be easily seen that the derived expression suggests a
chirality-like effect in the sense of its significant dependence
on mutual orientation of vectors $\vec k$, $\vec k_{0}$, and
$\hat{\vec{s}}^{\dag}\vec\sigma\hat{\vec{s}}$. Indeed, when
$\left[\vec k_{0},\vec
k\right]\bot\hat{\vec{s}}^{\dag}\vec\sigma\hat{\vec{s}}$, the
corresponding contribution to the fidelity is identically zero,
whereas at any other angle it is not zero and achieves a maximal
contribution to the fidelity when $\left[\vec k_{0},\vec
k\right]\|\hat{\vec{s}}^{\dag}\vec\sigma\hat{\vec{s}}$. In other
words, the statement that the error is small, which is understood
in the sense of the range from Eq. (\ref{fidelIrange}), determines
a relevant wave vector dependent range for the angle (we denote it
as $\Theta_{\vec k\hat{\vec{s}}}$) between $\left[\vec k_{0},\vec
k\right]$ and $\hat{\vec{s}}^{\dag}\vec\sigma\hat{\vec{s}}$.
Indeed, from Eqs. (\ref{fidelIrange}) and (\ref{fidelIII}) we have
\bea \label{anglerange}
\begin{array}{c}
\eta\leq A^{2}_{k_0}A^{2}_k\left(\left(1+B_{k_0}B_k\vec
k_{0}\cdot\vec k\right)^{2}\right.\\
\left.+\left(B_{k_0}B_k\left[\vec k_{0},\vec
k\right]\cdot\left(\hat{\vec{s}}^{\dag}\vec\sigma\hat{\vec{s}}\right)
\right)^{2}\right)\leq 1
\\
\displaystyle\Rightarrow
\frac{\left(\frac{\eta}{A^{2}_{k_0}A^{2}_k}-\left(1+B_{k_0}B_k\vec
k_{0}\cdot\vec
k\right)^{2}\right)}{\left(B_{k_0}B_k\right)^{2}\left|\left[\vec
k_{0},\vec k\right]\right|^{2}}\leq\cos^{2}\Theta_{\vec
k\hat{\vec{s}}}\\
\displaystyle\leq\frac{\left(\frac{1}{A^{2}_{k_0}A^{2}_k}-\left(1+B_{k_0}B_k\vec
k_{0}\cdot\vec
k\right)^{2}\right)}{\left(B_{k_0}B_k\right)^{2}\left|\left[\vec
k_{0},\vec k\right]\right|^{2}}.
\end{array}
\eea

Here, we used the fact that
$|\hat{\vec{s}}^{\dag}\vec\sigma\hat{\vec{s}}|=1$. Equation
(\ref{anglerange}) strictly establishes the relation between the
error $1-\eta$ and angle $\Theta_{\vec k\hat{\vec{s}}}$.

Let's evaluate how wide this range might be for narrow
noninteracting free electron beams when  $\left|\vec k-\vec k_0
\right| \ll\left|\vec k_0\right|$ is fulfilled. Here, it is
naturally to introduce a smallness parameter of the theory
$\chi=\frac{\left|\vec k-\vec k_0 \right|}{\left|\vec
k_0\right|}=\frac{\left|\Delta\vec k\right|}{\left|\vec
k_0\right|}\ll 1$.

Let's expand the both sides of the inequality from Eq.
(\ref{anglerange}) up to leading orders in $\chi$. It provides us
with the evaluation that we sought for. The expansion of the
left-hand side of Eq. (\ref{anglerange}) can be represented as
follows \bea \label{expleft}
\begin{array}{c}
\displaystyle\frac{\left(\frac{\eta}{A^{2}_{k_0}A^{2}_k}-\left(1+B_{k_0}B_k\vec
k_{0}\cdot\vec
k\right)^{2}\right)}{\left(B_{k_0}B_k\right)^{2}\left|\left[\vec
k_{0},\vec k\right]\right|^{2}}\\
\approx
C(\eta)_{-2}\chi^{-2}+C(\eta)_{-1}\chi^{-1}+C(\eta)_{0}\chi^{0}+...,
\end{array} \eea where \bea \label{expleftcoeff}
\begin{array}{c}
\displaystyle C(\eta)_{-2}=\frac{(\eta-1)}{e\sin^2\tau},\\
\displaystyle C(\eta)_{-1}=((\eta-1)f+\alpha(\eta))\frac{\vec k^{2}_{0}\cot\tau}{e\sin\tau},\\
\displaystyle C(\eta)_{0}=\left((\eta-1)g+\alpha(\eta)f
+\beta(\eta)\right)\frac{\vec k^{4}_{0}\cot^{2}\tau}{e},
\end{array}
\eea with \bea \label{expleftcoeffparamI}
e=(A_{k_0}B_{k_0})^{4}\vec k^{4}_{0}, \quad
f=\frac{2\Delta_B}{B_{k_0}}, \quad
g=\frac{\Delta^{2}_B}{B^{2}_{k_0}}, \eea \bea
\label{expleftcoeffparamII}
\begin{array}{c}
\displaystyle {\alpha(\eta)=\frac{2\eta\Delta_A}{A_{k_0}}-2A^{2}_{k_0}B^{2}_{k_0}+2A^{2}_{k_0}B_{k_0}\Delta_B\vec k^{2}_{0}},\\
\displaystyle \beta(\eta)=\frac{\eta\Delta^{2}_{A}}{A^{2}_{k_0}}+2A^{2}_{k_0}B_{k_0}\Delta_B-A^{4}_{k_0}B^{4}_{k_0}\\
\displaystyle+2A^{4}_{k_0}B^{3}_{k_0}\Delta_B\vec
k^{2}_{0}-A^{4}_{k_0}B^{2}_{k_0}\Delta^{2}_B\vec k^{4}_{0},
\end{array}
\eea \bea \label{expleftcoeffparamIII}
\begin{array}{c}
A_{k}\approx A_{k_0}-\Delta_{A}\vec k_{0}\Delta\vec k=A_{k_0}-\chi\Delta_{A}\vec k^{2}_{0}\cos\tau,\\
B_{k}\approx B_{k_0}-\Delta_{B}\vec k_{0}\Delta\vec
k=B_{k_0}-\chi\Delta_{B}\vec k^{2}_{0}\cos\tau,
\end{array}
\eea \bea \label{expleftcoeffparamIV}
\begin{array}{c}
\displaystyle \Delta_{A} = \frac{\hbar^{2}m_e
c^{4}}{4E^{3}_{k_0}},\quad
\Delta_{B} = \frac{\hbar^{3}c^{3}}{E_{k_0}(E_{k_0}+m_e c^{2})^{2}},\\
 \displaystyle E_{k} \approx  E_{k_0}+\frac{\hbar^{2}c^{2}}{E_{k_0}}\vec k_{0}\Delta\vec k = E_{k_0}+\chi\frac{\hbar^{2}c^{2}}{E_{k_0}}\vec k^{2}_{0}\cos\tau.
\end{array}
\eea

Here, $\tau$ is an angle between $\vec k_{0}$ and $\Delta\vec k$.
Then, consequently, for the right- hand side of Eq.
(\ref{anglerange}) we come to the expansion \bea \label{expright}
\begin{array}{c}
\displaystyle
\frac{\left(\frac{1}{A^{2}_{k_0}A^{2}_k}-\left(1+B_{k_0}B_k\vec
k_{0}\cdot\vec
k\right)^{2}\right)}{\left(B_{k_0}B_k\right)^{2}\left|\left[\vec
k_{0},\vec k\right]\right|^{2}}\\
\approx C(\eta=1)_{-1}\chi^{-1}+C(\eta=1)_{0}\chi^{0}+...,
\end{array}\eea since, $C(\eta=1)_{-2}\equiv 0$, according to Eq.
(\ref{expleftcoeff}). Eventually, based on Eqs.
(\ref{anglerange})-(\ref{expright}), the evaluation for the width
of the range for  $\cos^{2}\Theta_{\vec k\hat{\vec{s}}}$ (we
denote it as $\Delta_{\cos^{2}\Theta_{\vec k\hat{\vec{s}}}}$) can
be written as follows \bea \label{width}
\begin{array}{c}
\displaystyle \Delta_{\cos^{2}\Theta_{\vec
k\hat{\vec{s}}}}=\frac{\left(\frac{1}{A^{2}_{k_0}A^{2}_k}-\left(1+B_{k_0}B_k\vec
k_{0}\cdot\vec k\right)^{2}\right)}
{\left(B_{k_0}B_k\right)^{2}\left|\left[\vec k_{0},\vec
k\right]\right|^{2}}\\
\displaystyle -\frac{\left(\frac{\eta}{A^{2}_{k_0}A^{2}_k}-\left(1+B_{k_0}B_k\vec k_{0}\cdot\vec k\right)^{2}\right)}{\left(B_{k_0}B_k\right)^{2}\left|\left[\vec k_{0},\vec k\right]\right|^{2}}\\
\approx-C(\eta)_{-2}\chi^{-2}+ (C(\eta=1)_{-1}-C(\eta)_{-1})\chi^{-1}\\
+(C(\eta=1)_{0}-C(\eta)_{0})\chi^{0}
\displaystyle =\frac{(1-\eta)}{e\sin^2\tau}\chi^{-2}\\
\displaystyle + ((1-\eta)f+\alpha(\eta=1)-\alpha(\eta))\frac{\vec k^{2}_{0}\cot\tau}{e\sin\tau}\chi^{-1}\\
+\left((1-\eta)g+(\alpha(\eta=1)-\alpha(\eta))f\right.\\
\left. +\beta(\eta=1)-\beta(\eta)\right)\displaystyle \frac{\vec k^{4}_{0}\cot^{2}\tau}{e}\chi^{0}\\
\displaystyle=(1-\eta)\left(\frac{1}{e\sin^2\tau}\chi^{-2}+(f+\frac{2\Delta_A}{A_{k_0}})\frac{\vec k^{2}_{0}\cot\tau}{e\sin\tau}\chi^{-1}\right.\\
\displaystyle\left.+\left(g+\frac{2\Delta_A}{A_{k_0}}f+\frac{\Delta^{2}_{A}}{A^{2}_{k_0}}\right)\frac{\vec k^{4}_{0}\cot^{2}\tau}{e}\chi^{0}\right)\\
\displaystyle=\frac{(1-\eta)}{(A_{k_0}B_{k_0})^{4}\vec k^{4}_{0}}
\\
\displaystyle \times\left(\frac{1}{\sin^2\tau}\chi^{-2}+2(\frac{\Delta_B}{B_{k_0}}+\frac{\Delta_A}{A_{k_0}})\frac{\vec k^{2}_{0}\cot\tau}{\sin\tau}\chi^{-1}\right.\\
\displaystyle \left.+\left(\frac{\Delta^{2}_B}{B^{2}_{k_0}}+
\frac{4\Delta_A\Delta_B}{A_{k_0}B_{k_0}}+\frac{\Delta^{2}_{A}}{A^{2}_{k_0}}\right)\vec
k^{4}_{0}\cot^{2}\tau\chi^{0}\right)\leq 1.
\end{array}
\eea

Here, trigonometry dictates the last line inequality.

It can be immediately seen that there is a very particular
configuration when $\cos\tau=0\Rightarrow\cot\tau\equiv 0$,
geometrically corresponding to those $\vec k$ for which $\vec
k_{0}\bot\Delta\vec k$. Indeed, since all terms are positive, we
are expecting a minimum value for the range that takes the
following analytical form \bea \label{widthmin}
\Delta^{min}_{\cos^{2}\Theta_{\vec
k\hat{\vec{s}}}}=\frac{(1-\eta)\chi^{-2}}{(A_{k_0}B_{k_0})^{4}\vec
k^{4}_{0}}. \eea

In addition, it gives us a hint for an optimal configuration that
provides a maximal fidelity, i.e., the lowest error, within this
approximation. Namely, if, along with $\vec k_{0}\bot\Delta\vec
k$, we require that $[\vec k_{0},\Delta\vec
k]\bot\hat{\vec{s}}^{\dag}\vec\sigma\hat{\vec{s}}$, then from Eqs.
(\ref{fidelIII}) and (\ref{expleftcoeffparamIII}) we have
\bea\label{fidelorth}
\begin{array}{c}
F=A^{2}_{k_0}A^{2}_{k}\left(\left(1+B_{k_0}B_k\vec k_{0}\cdot\left(\vec k_0+\Delta\vec k\right)\right)^{2}\right.\\
\left.+\left(B_{k_0}B_k\left[\vec k_{0},\left(\vec k_0+\Delta\vec k\right)\right]\cdot\left(\hat{\vec{s}}^{\dag}\vec\sigma\hat{\vec{s}}\right) \right)^{2}\right)\\
=A^{2}_{k_0}A^{2}_{k}\left(1+B_{k_0}B_k\vec
k^{2}_{0}\right)^{2}\approx A^{4}_{k_0}\left(1+B^{2}_{k_0}\vec
k^{2}_{0}\right)^{2}\equiv 1.
\end{array}
\eea

Moreover, since, based on Eqs. (\ref{width}) and (\ref{widthmin}),
\bea \label{widthminineq} \Delta^{min}_{\cos^{2}\Theta_{\vec
k\hat{\vec{s}}}}\leq 1, \eea then, the following error constraint
takes place \bea \label{errorconst} (1-\eta)\leq
(A_{k_0}B_{k_0})^{4}\vec k^{4}_{0}\chi^{2}. \eea
\section{Conclusion}
Thus, we can conclude that: first, the quantum information
analysis demonstrates that the error evaluation inevitably
requires taking into consideration the third vector
$\hat{\vec{s}}^{\dag}\vec\sigma\hat{\vec{s}}$ along with two
commonly used $\vec k$ and $\vec k_{0}$, leading to a
chirality-like effect in the error, since, the fidelity is
extremely sensitive to the mutual configuration of these three
vectors, and a relevant range of characteristic angular dependent
quantity $\cos^{2}\Theta_{\vec k\hat{\vec{s}}}$ is analytically
established; second, within the $\left|\vec k-\vec k_0 \right|
\ll\left|\vec k_0\right|$  approximation, the least, say zero,
error  is achieved for spinors with those wave vectors and spins
for which $\vec k_0$, $\Delta\vec k$, and
$\hat{\vec{s}}^{\dag}\vec\sigma\hat{\vec{s}}$ forms a set of three
mutually orthogonal vectors, and, at fixed $\vec k_0$ and
$\hat{\vec{s}}^{\dag}\vec\sigma\hat{\vec{s}}$, that defines a
zero-error line for the ends of $\vec k$; third, within the same
approximation a minimal range width of valid $\cos^{2}\Theta_{\vec
k\hat{\vec{s}}}$ is evaluated and turned out to be defined by both
the error $1-\eta$ and the approximation smallness parameter
$\chi$; fourth, trigonometry allows us to establish a direct
constraint on the error, which quadratically depends on $\chi$.

\emph{Data availability}--No datasets were generated or analysed
during the current study.

%\appendix
%\section{}

%\begin{acknowledgments}
%\end{acknowledgments}

\bibliography{bibliographynew.bib}

@article{Abajo2019,
 	author = "Valerio Di Giulio and Mathieu Kociak and F. Javier Garc\'{i}a de Abajo",
	journal = "Optica",
	number = "12",
	pages = "1524--1534",
	title = "Probing quantum optical excitations with fast electrons",
	volume = "6",
	month = "Dec",
	year = "2019",
	doi = "10.1364/OPTICA.6.001524"
}

@article{Velasco2024,
	author = "Velasco, Cruz I. and Di Giulio, Valerio and  de Abajo, F. Javier Garc\'{i}a",
	title = "Radiative loss of coherence in free electrons: a long-range quantum phenomenon",
	journal = "Light {S}ci. {A}ppl.",
	volume = "13",
	pages = "31",
	year = "2024",
	doi = "10.1038/s41377-023-01361-6"
}

@misc{Velasco2026,
  	author="Velasco, Cruz I. and Di Giulio, Valerio and  de Abajo, F. Javier Garc\'{i}a",
	title="Free-electron decoherence: Theory and applications",
 	year="2026",
	archivePrefix={arXiv},
	eprint={2602.14693},
	url={https://arxiv.org/abs/2602.14693}
}

@book{Nielsen2010, 
	author="Nielsen, M. A. and Chuang, I. L.",
	title="Quantum {C}omputation and {Q}uantum {I}nformation", 
	address="Cambridge",
	publisher="Cambridge {U}niversity {P}ress",
	year="2010"	
 }

@book{Sakurai, 
	author="Sakurai, Jun John and Napolitano, Jim", 
	title="Modern {Q}uantum {M}echanics", 
	address="Cambridge",
	publisher="Cambridge {U}niversity {P}ress", 
	year="2020"	
}
\end{document}